\documentclass[aps,prd,showkeys,superscriptaddress,twocolumn,nofootinbib,floatfix]{revtex4-1}

\usepackage{amsmath}
\usepackage{amssymb}
\usepackage{amsthm}
\usepackage{mathrsfs}
\usepackage{graphicx}
\usepackage{epstopdf}
\usepackage{fancyhdr}
\usepackage{array}
\usepackage[all]{xy}
\usepackage{eufrak}
\usepackage{euscript}
\usepackage{enumerate}
\usepackage{slashed}
\usepackage{hyperref}
\usepackage{subfigure} 
\usepackage{epstopdf} 

\hypersetup{pdftex,colorlinks=true,linkcolor=blue,citecolor=blue,menucolor=black,urlcolor=blue,filecolor=blue}

\begin{document}

\title{Chern-Simons diffusion rate across different phase transitions}

\author{Romulo Rougemont}
\email{romulo@if.usp.br}
\affiliation{Instituto de F\'{i}sica, Universidade de S\~{a}o Paulo, Rua do Mat\~{a}o, 1371, Butant\~{a}, CEP 05508-090, S\~{a}o Paulo, SP, Brazil}

\author{Stefano Ivo Finazzo}
\email{stefano@ift.unesp.br}
\affiliation{Instituto de F\'{i}sica Te\'orica, Universidade do Estado de S\~{a}o Paulo, Rua Dr. Bento T. Ferraz, 271, CEP 01140-070, S\~{a}o Paulo, SP, Brazil}

\begin{abstract}
We investigate how the dimensionless ratio given by the Chern-Simons diffusion rate $\Gamma_{\textrm{CS}}$ divided by the product of the entropy density $s$ and temperature $T$ behaves across different kinds of phase transitions in the class of bottom-up non-conformal Einstein-dilaton holographic models originally proposed by Gubser and Nellore. By tuning the dilaton potential, one is able to holographically mimic a first order, a second order, or a crossover transition. In a first order phase transition, $\Gamma_{\textrm{CS}}/sT$ jumps at the critical temperature (as previously found in the holographic literature), while in a second order phase transition it develops an infinite slope. On the other hand, in a crossover, $\Gamma_{\textrm{CS}}/sT$ behaves smoothly, although displaying a fast variation around the pseudo-critical temperature. In all the cases, $\Gamma_{\textrm{CS}}/sT$ increases with decreasing $T$. The behavior of the Chern-Simons diffusion rate across different phase transitions is expected to play a relevant role for the chiral magnetic effect around the QCD critical end point, which is a second order phase transition point connecting a crossover band to a line of first order phase transition.  Our findings in the present work add to the literature the first predictions for the Chern-Simons diffusion rate across second order and crossover transitions in strongly coupled non-conformal, non-Abelian gauge theories.
\end{abstract}


\keywords{Chern-Simons diffusion rate, finite temperature, holography.}

\maketitle

\section{Introduction}

The vacuum of non-Abelian gauge theories defined in Euclidean space possesses a rich topological structure featuring, for instance, non-trivial local minima ($F_{\mu\nu}^a\neq 0$) distinct from the trivial global minimum of the theory. These local minima are labeled by the values of the Pontryagin index, Chern-Simons number or topological charge of the associated gauge field configurations and correspond to different topological sectors (or ``vacua'') of the true, gauge-invariant vacuum of the system, the so-called $\theta$-vacuum, which is given by a superposition of the different topological vacua of the theory. Within each topological sector, the action of the system is bounded from below by a multiple of the (absolute value of the) topological charge, with the bound being saturated by topological field configurations satisfying the self-duality condition $F_{\mu\nu}^a=\pm\tilde{F}_{\mu\nu}^a\equiv\pm\frac{1}{2}\epsilon_{\mu\nu\alpha\beta}F^{\alpha\beta\,a}$ \cite{Jackiw:1976fs}. Such topological configurations at zero temperature are called \emph{instantons} \cite{Belavin:1975fg,'tHooft:1976fv}. An instanton with topological charge $Q$ may be interpreted as a quantum tunneling process between two different topological vacua labeled by $|N\rangle$ and $|N+Q\rangle$. At nonzero temperature, thermal saddle point configurations of the non-Abelian gauge field which are able to modify the topological charge may be also relevant. In the context of the electroweak sector of the standard model, these thermal configurations are called \emph{sphalerons} \cite{Manton:1983nd,Klinkhamer:1984di}.

The Chern-Simons diffusion rate $\Gamma_{\textrm{CS}}$ is a real time, non-equilibrium observable measuring the rate of change of the topological charge per unit of volume and per unit of time. This observable has vast phenomenological relevance. For instance, the Chern-Simons diffusion rate in the electroweak sector is related to the violation of baryon number and, therefore, to the baryogenesis process in the early universe \cite{Kuzmin:1985mm,Arnold:1987mh,Cohen:1991iu,Cohen:1993nk}. In the context of the strong interaction, the Chern-Simons diffusion rate is relevant to the study of the Chiral Magnetic Effect (CME) expected to take place in non-central heavy ion collisions \cite{Kharzeev:2004ey,Kharzeev:2007tn,Kharzeev:2007jp,Fukushima:2008xe,Kharzeev:2015kna,Kharzeev:2015znc}.

In the perturbative regime of Yang-Mills theory, valid at weak coupling, the Chern-Simons diffusion rate has been widely studied in the literature, see for instance Refs. \cite{Arnold:1996dy,Moore:1997sn,Moore:1999fs,Bodeker:1999gx,Moore:2010jd} and references therein. In order to access the strong coupling regime of some non-Abelian gauge theories with gravity duals, the gauge/gravity duality \cite{Maldacena:1997re,Witten:1998qj,Witten:1998zw,Gubser:1998bc} has been employed to compute the Chern-Simons diffusion rate in a large variety of different holographic settings, as in Refs. \cite{Son:2002sd,Basar:2012gh,Craps:2012hd,Gursoy:2012bt,Jahnke:2014sla,Drwenski:2015sha}. However, no systematic study on the general properties of the Chern-Simons diffusion rate across different types of phase transitions has been undertaken on the literature yet.

The Beam Energy Scan (BES) program at the Relativistic Heavy Ion Collider (RHIC) is currently probing regions of the QCD phase diagram at finite temperature $T$ and nonzero baryon chemical potential $\mu_B$ where the quark-gluon plasma (QGP) \cite{Gyulassy:2004zy,Heinz:2013th,Shuryak:2014zxa} produced in ultrarelativistic heavy ion collisions \cite{Adcox:2004mh,Arsene:2004fa,Back:2004je,Adams:2005dq} is highly non-conformal and behaves as a strongly coupled fluid. One of the main aims of the BES is to find out the QCD critical end point (CEP) location in the $(T,\mu_B)$ phase diagram, which is a second order phase transition point connecting the crossover band at lower $\mu_B$ to a line of first order phase transition at higher $\mu_B$. Therefore, around the QCD CEP, one expects the system to probe different kinds of phase transitions and the behavior of the Chern-Simons diffusion rate in this region is of particular relevance for the phenomenology of the CME. As a first \textit{qualitative glimpse} into this issue, one might study the behavior of the Chern-Simons diffusion rate in simple models displaying different kinds of phase transitions.

In this work, we employ the class of bottom-up non-conformal Einstein-dilaton holographic models originally proposed by Gubser and Nellore in Ref. \cite{Gubser:2008ny} and investigate how the dimensionless ratio $\Gamma_{\textrm{CS}}/sT$ behaves across different types of phase transitions. By adequately adjusting the dilaton potential one is able to induce, on the dual gauge field theory, a first order, a second order, or a crossover transition. As originally found in Ref. \cite{Gursoy:2012bt}, we conclude that in a first order phase transition $\Gamma_{\textrm{CS}}/sT$ jumps at the critical temperature. As the new results of the present work, we find that in a second order phase transition $\Gamma_{\textrm{CS}}/sT$ develops an infinite slope at the critical temperature, while displaying a smooth but fast varying temperature profile around the pseudo-critical temperature in a crossover transition. In all the cases, $\Gamma_{\textrm{CS}}/sT$ increases with decreasing $T$. Our conclusions are shown to be robust in face of the choice of different profiles for the axion-dilaton coupling previously derived in the literature.

We use natural units where $\hbar = k_B = c = 1$ and a mostly plus metric signature.

\section{Einstein-dilaton holographic model}

In this section, we briefly review the main results concerning the Einstein-dilaton holographic setup originally proposed in Ref. \cite{Gubser:2008ny}, which has been also extensively discussed, for instance, in Refs. \cite{Gubser:2008yx,Finazzo:2014zga,Yaresko:2015ysa}. We refer the reader to consult these references for further details.

The bulk action is given by
\begin{align}
S=\frac{1}{2\kappa^2} \int d^5x \, \sqrt{-g} \left[R-\frac{1}{2}(\partial_\mu \phi)^2-V(\phi)\right],
\label{eq:EDaction}
\end{align}
while the black hole ansatze for the metric and the dilaton field describing a plasma at the finite temperature gauge field theory may be written as follows
\begin{align}
ds^2=\frac{e^{B(r)}\,dr^2}{h(r)}+e^{2A(r)}\left[-h(r)dt^2+d\vec{x}^2\right],\,\,\,\phi=\phi(r).
\label{eq:ansatze}
\end{align}
In this work, we fix the gauge corresponding to the condition $r=\phi$, which may be attained due to the freedom of reparameterization of the radial coordinate. The radial position of the black hole horizon is given by the largest root of $h(\phi_H)=0$, while the boundary of the asymptotically AdS$_5$ space lies at $\phi=0$.

The functional form of the dilaton potential we shall employ in the present work is given by
\begin{align}
V(\phi)=\frac{-12\,\left(1+a\phi^2\right)^{1/4}\,\cosh(\gamma\phi)+b_2\phi^2+b_4\phi^4+b_6\phi^6}{L^2}.
\label{eq:Vphi}
\end{align}
In the present work, we set to unity the radius $L$ of the asymptotically AdS$_5$ space. Different choices of the free parameters $a$, $\gamma$, $b_2$, $b_4$, and $b_6$ in Eq. \eqref{eq:Vphi} may mimic the properties of physical systems with different kinds of phase transitions in the gauge theory \cite{Gubser:2008ny,Gubser:2008yx,Finazzo:2014zga}, and we shall explore it in the next sections in order to investigate the behavior of the Chern-Simons diffusion rate in first and second order phase transitions and also in a crossover.

As discussed in Ref. \cite{Gubser:2008ny}, the Einstein-dilaton equations of motion in the $r=\phi$ gauge may be cast into the form of a single master equation, namely,
\begin{align}
\frac{G'}{G+V/3V'}=\frac{d}{d\phi}\ln\left(\frac{6G'+1-24G^2}{6G}-\frac{G'}{G+V/3V'}\right),
\label{eq:master}
\end{align}
where we defined the master function $G(\phi)\equiv A'(\phi)$. The master equation \eqref{eq:master} may be numerically solved by integrating it out from the horizon $\phi=\phi_H$ up to the boundary $\phi=0$ with the following initial conditions imposed at the horizon,
\begin{align}
G(\phi_H)&=-\frac{V(\phi_H)}{3V'(\phi_H)}, \\
G'(\phi_H) &= \frac{1}{6} \left( \frac{V(\phi_H) V''(\phi_H)}{V'(\phi_H)^2}  -1 \right),
\label{eq:initialcond}
\end{align} 
which follow from Einstein's equations \cite{Finazzo:2014zga}. Each value of $\phi_H$ translates into some physical state in the gauge theory with definite values of temperature $T$ and entropy density $s$.

The full background geometry can be reconstructed from the master function $G(\phi)$ using the following relations \cite{Gubser:2008ny,Finazzo:2014zga}:
\begin{align}
A(\phi,\phi_H)&=\frac{\ln(\phi_H)}{\Delta-4}+\int_0^{\phi_H}d\phi'\left[G(\phi')-\frac{1}{(\Delta-4)\phi'}\right]\nonumber\\
& +\int_{\phi_H}^{\phi}d\phi' G(\phi'),\label{eq:Aphi}\\
B(\phi,\phi_H)&=\ln\left(\frac{V(\phi_H)}{3V'(\phi_H)}\right)+\int_0^{\phi_H}\frac{d\phi'}{6G(\phi')}\nonumber\\
&+\int_{\phi_H}^{\phi}d\phi' \frac{6G'(\phi')+1}{6G(\phi')},\label{eq:Bphi}\\
h(\phi,\phi_H)&=\frac{\int_{\phi_H}^{\phi}d\phi' e^{B(\phi')-4A(\phi')}}{\int_{\phi_H}^0 d\phi' e^{B(\phi')-4A(\phi')}},\label{eq:hphi}
\end{align}
where $\Delta$ is the scaling dimension of the gauge field theory operator dual to the bulk dilaton field, which is given by the largest root of $\Delta(\Delta-4)=m^2$, where $m$ is the dilaton mass extracted from the potential \eqref{eq:Vphi}. We remark that for all the potentials used in the next sections, the Breitenlohner-Freedman bound \cite{Breitenlohner:1982jf,Breitenlohner:1982bm} is satisfied. The temperature and entropy density of a given black hole solution read, respectively,
\begin{align}
T(\phi_H)&=\frac{e^{A(\phi_H)-B(\phi_H)}|h'(\phi_H)|}{4\pi},\label{eq:TphiH}\\
s(\phi_H)&=\frac{2\pi e^{3A(\phi_H)}}{\kappa^2}.\label{eq:sphiH}
\end{align}

\section{Axion field and Chern-Simons diffusion rate}

In this section, we review the derivation of the holographic formula for the Chern-Simons diffusion rate in general dilaton black hole backgrounds originally obtained in Ref. \cite{Gursoy:2012bt}. This formula holds when the black hole solutions are the dominant saddle points of the system, and as we are going to discuss in the next sections, this will be always the case for the potentials used here emulating a second order phase transition or a crossover, while for the case of a first order phase transition it will hold only for $T\ge T_c$, where $T_c$ is the critical temperature of the phase transition.

The Chern-Simons diffusion rate is a real time observable which measures the rate of change of the topological charge per unit of volume and per unit of time. It may be calculated through the following Kubo's formula \cite{Gursoy:2012bt},
\begin{align}
\Gamma_{\textrm{CS}}=-\lim_{\omega\rightarrow 0}\frac{2T}{\omega}\,\textrm{Im}\left[G_{qq}^{(R)}(\omega,\vec{k}=\vec{0})\right],
\label{eq:Gammadef}
\end{align}
where $G_{qq}^{(R)}$ is the thermal retarded propagator of $q(x)=(g^2/64\pi^2)\epsilon_{\mu\nu\alpha\beta}F^{\mu\nu}_a(x)F^{\alpha\beta}_a(x)$, the instanton number density operator sourced in the gauge theory by the $\theta$-angle.\footnote{The topological charge $Q$ is related to the instanton number density operator $q(x)$ by the expression $Q\equiv\int d^4x\, q(x)$.} According to the holographic dictionary, the $\theta$-angle is given by the boundary value of a bulk axion field in an asymptotically AdS$_5$ space.

We work here in the classical gravity approximation of the holographic correspondence, valid at strong coupling and large number of colors $N_c$ in the gauge theory. In this regime, the axion field is taken as a probe over the background solutions of the Einstein-dilaton model, with its action given by\footnote{The axion is a probe in the large $N_c$ limit because its action is $N_c^2$ suppressed relatively to the Einstein-dilaton action. As so, one could formally write a factor of $N_c^{-2}$ in front of Eq. \eqref{eq:axionaction}, as done, for instance, in Ref. \cite{Gursoy:2012bt}. Here, we absorb this factor into the overall normalization of the axion-dilaton coupling funtion $Z(\phi)$.}
\begin{align}
S_{\textrm{axion}}=\frac{1}{2\kappa^2} \int d^5x \, \sqrt{-g} \left[-\frac{Z(\phi)}{2}(\partial_\mu \alpha)^2\right],
\label{eq:axionaction}
\end{align}
where $Z(\phi)$ is the axion-dilaton coupling and $\alpha(\phi=0)=\theta$. Then, through the holographic correspondence, correlation functions of $q(x)$ in the gauge theory are extracted from the on-shell action for the axion field.
More precisely, in order to pursue the holographic calculation of the thermal retarded two-point Green's function in Eq. \eqref{eq:Gammadef}, one follows the prescription originally introduced in Ref. \cite{Son:2002sd}:
\begin{enumerate}
\item The first step is to consider a plane-wave ansatz for the Fourier modes of a probe axion fluctuation on top of the (Einstein-dilaton) backgrounds;\footnote{Since we are working in the probe approximation for the axion field, its background value is zero.}
\item Next, one substitutes this plane-wave ansatz into the probe action \eqref{eq:axionaction} and obtains the linearized equation of motion for the axion perturbation, which must be subjected to the boundary condition that the perturbation is normalized to unity at the boundary and that the wave is in-falling at the horizon;
\item Then, in order to obtain the retarded propagator in Eq. \eqref{eq:Gammadef}, one needs to solve the equation of motion for the perturbation with $\vec{k}=\vec{0}$ and small $\omega$, plug the solution into the action \eqref{eq:axionaction} and discard the horizon piece of the on-shell disturbed action, working solely with the boundary on-shell action.
\end{enumerate}

The linearized equation of motion for the probe axion perturbation $\delta\alpha$ reads
\begin{align}
&\partial_{\phi}\left[Z(\phi)\sqrt{-g}g^{\phi\phi}\partial_{\phi}\delta\alpha(\phi,x^\mu)\right]\nonumber\\ &+Z(\phi)\sqrt{-g}g^{\mu\nu}\partial_\mu\partial_\nu\delta\alpha(\phi,x^\mu)=0,
\label{eq:axioneomrealspace}
\end{align}
where
\begin{align}
\delta\alpha(\phi,x^\mu)=\int\frac{d^4k}{(2\pi)^4} e^{ik_\mu x^\mu} \delta\alpha(\phi,k^\mu) a(k^\mu),
\label{eq:fourier}
\end{align}
with $k_\mu=(-\omega,\vec{k})$ and $a(k^\mu)$ is the source term for the gauge theory operator $q(x)$. At the boundary we fix the Dirichlet condition:
\begin{align}
\lim_{\phi\rightarrow 0}\delta\alpha(\phi,k^\mu)=1.
\label{eq:Dirichlet}
\end{align}
Eq. \eqref{eq:axioneomrealspace} reads in the Fourier space,
\begin{align}
&\partial_{\phi}\left[Z(\phi)\sqrt{-g}g^{\phi\phi}\partial_{\phi}\delta\alpha(\phi,x^\mu)\right]\nonumber\\ &-Z(\phi)\sqrt{-g}g^{\mu\nu}k_\mu k_\nu\delta\alpha(\phi,x^\mu)=0,
\label{eq:axioneom}
\end{align}
which must be solved with the Dirichlet condition \eqref{eq:Dirichlet} at the boundary and with the in-falling wave condition at the horizon.

The boundary on-shell action may be written in the form
\begin{align}
S_{\textrm{axion}}^{\textrm{bdy}}=-\frac{1}{2}\int\frac{d^4k}{(2\pi)^4}a(-k^\mu)G_{qq}^{(R)}(\omega,\vec{k})a(k^\mu),
\label{eq:on-shell}
\end{align}
where
\begin{align}
G_{qq}^{(R)}(\omega,\vec{k})=\frac{1}{2\kappa^2}\lim_{\phi\rightarrow 0}Z(\phi)\sqrt{-g}g^{\phi\phi}\partial_{\phi} \delta\alpha(\phi,k^\mu).
\label{eq:propagator}
\end{align}

Due to the Kubo's formula \eqref{eq:Gammadef}, in order to obtain the Chern-Simons diffusion rate we just need to evaluate \eqref{eq:propagator} at $\vec{k}=\vec{0}$ and $\omega\ll T$: in this case, as originally discussed in Ref. \cite{Gursoy:2012bt}, it is simple to solve Eq. \eqref{eq:axioneom} using the near-horizon matching technique, or, alternatively, the membrane paradigm \cite{Iqbal:2008by}. We employ here the near-horizon matching method, which consists in first solving Eq. \eqref{eq:axioneom} with $\omega=0$ and then expanding the result around the horizon, with the outcome of such calculation being matched with the outcome of the reverse order of operations, where one first solves the asymptotic form of Eq. \eqref{eq:axioneom} near the horizon and then expands the result around $\omega=0$.

The first step is to solve Eq. \eqref{eq:axioneom} with $\vec{k}=\vec{0}$ and $\omega=0$, which may be done in closed form in terms of the background:
\begin{align}
\delta\alpha\biggr|_{\omega=0}=C_1+C_2\int_0^{\phi}\frac{d\phi'}{Z(\phi')e^{4A(\phi')-B(\phi')}h(\phi')}.
\label{eq:solution1}
\end{align}
By applying the Dirichlet condition \eqref{eq:Dirichlet} into Eq. \eqref{eq:solution1}, one fixes $C_1=1$, and substituting it into Eq. \eqref{eq:propagator}, one finds that
\begin{align}
G_{qq}^{(R)}(\omega\ll T,\vec{k}=\vec{0})=\frac{C_2}{2\kappa^2}.
\label{eq:propagator1}
\end{align}

In order to fix the integration constant $C_2$, we expand \eqref{eq:solution1} around $\phi=\phi_H$, using that $h(\phi\rightarrow\phi_H)=-h'(\phi_H)(\phi_H-\phi)+\mathcal{O}(\phi_H-\phi)^2$,
\begin{align}
\delta\alpha\biggr|_{\omega=0}^{\phi\rightarrow\phi_H}\approx 1+\frac{C_2}{4\pi TZ(\phi_H)e^{3A(\phi_H)}} \ln(\phi_H-\phi),
\label{eq:solution2}
\end{align}
where we used Eq. \eqref{eq:TphiH}.

Now we revert the order of operations and first solve the asymptotic form of Eq. \eqref{eq:axioneom} with $\vec{k}=\vec{0}$ close to the horizon, obtaining
\begin{align}
\delta\alpha\biggr|_{\phi\rightarrow\phi_H}=C_-(\phi_H-\phi)^{-i\omega/4\pi T}+C_+(\phi_H-\phi)^{i\omega/4\pi T}.
\label{eq:solution3}
\end{align}
The in-falling wave condition at the horizon is imposed by setting $C_+=0$. Now we expand the result around $\omega=0$,
\begin{align}
\delta\alpha\biggr|_{\phi\rightarrow\phi_H}^{\omega\rightarrow 0}=C_-\left[1-\frac{i\omega}{4\pi T}\ln(\phi_H-\phi)+ \mathcal{O}\left(\frac{\omega}{4\pi T}\right)^2\right].
\label{eq:solution4}
\end{align}

Finally, we match the terms of same order in Eqs. \eqref{eq:solution2} and \eqref{eq:solution4}, fixing the integration constants
\begin{align}
C_-=1,\,\,\, C_2=-i\omega Z(\phi_H)e^{3A(\phi_H)}.
\label{eq:constants}
\end{align}
Then, by using Eq. \eqref{eq:sphiH}, one finds that \cite{Gursoy:2012bt}
\begin{align}
\frac{\Gamma_{\textrm{CS}}}{sT}=\frac{Z(\phi_H)}{2\pi},
\label{eq:GammaCS}
\end{align}
where the temperature dependence of the dimensionless ratio $\Gamma_{\textrm{CS}}/sT$ is encoded in the interplay between the form of the axion-dilaton coupling function $Z(\phi)$ and the inverse relation $\phi_H(T)$, with the latter being determined by the form of the dilaton potential $V(\phi)$. In the following sections, we are going to discuss how different choices of $V(\phi)$ translate in terms of the behavior of $\Gamma_{\textrm{CS}}/sT$ across different kinds of phase transitions and will also discuss the influence of different choices of $Z(\phi)$ on the results.

The different potentials we shall employ in the present work fall within the functional form of Eq. \eqref{eq:Vphi}. Regarding the axion-dilaton coupling, we shall consider here two different functional forms, namely,
\begin{align}
Z(\phi)&=Z_0(1+Z_1e^{\sqrt{6}\phi}),\,\,\,\textrm{or}\label{eq:ZKiritsis}\\
Z(\phi)&=e^{2\phi}.\label{eq:ZTrancanelli}
\end{align}
Since we shall present our results for $\Gamma_{\textrm{CS}}/sT$ normalized by its value at $T\rightarrow\infty$, in face of Eq. \eqref{eq:GammaCS}, the free parameter $Z_0$ in the profile \eqref{eq:ZKiritsis} drops out in such normalization and we just need to consider different values of $Z_1$. The axion-dilaton profile in Eq. \eqref{eq:ZKiritsis} was proposed in Ref. \cite{Gursoy:2012bt} in the context of a bottom-up Einstein-dilaton holographic model mimicking the first order phase transition seen on lattice simulations of pure Yang-Mills theory\footnote{The factor of $\sqrt{3/8}$ in $\sqrt{6}=\sqrt{3/8}\times 4$ comes from the different normalizations for the kinetic term of the dilaton field employed here and in Ref. \cite{Gursoy:2012bt}.}, while the profile in Eq. \eqref{eq:ZTrancanelli} was obtained in Refs. \cite{Mateos:2011ix,Mateos:2011tv} in a top-down Einstein-axion-dilaton setup constructed as a solution of type IIB supergravity dual to a spatially anisotropic $\mathcal{N}=4$ Super Yang-Mills plasma. As we are going to show in the next sections, both profiles in Eqs. \eqref{eq:ZKiritsis} and \eqref{eq:ZTrancanelli} produce qualitatively similar results for $\Gamma_{\textrm{CS}}/sT$. The distinctive qualitative features of $\Gamma_{\textrm{CS}}/sT$ we shall see in the following sections will be determined by the nature of the phase transition considered, which in turn depends on the chosen parameters of the dilaton potential in Eq. \eqref{eq:Vphi}.

\section{Model A: first order phase transition}

\begin{table}[h]
 \begin{center}
  \begin{tabular}{| c | c | c | c | c |}
    \hline
    $a$ & $\gamma$ & $b_2$ & $b_4$ & $b_6$ \\
    \hline
    \hline
    1 & $\sqrt{2/3}$ & $5.5$ & $0.3957$ & $0.0135$ \\
    \hline
  \end{tabular}
 \caption{A choice of parameters of the dilaton potential in Eq. \eqref{eq:Vphi} yielding a system with a first order phase transition. \label{tabA}}
 \end{center}
\end{table}

In this section, we discuss the thermodynamics and the behavior of the Chern-Simons diffusion rate in a system with a first order phase transition, which may be modeled by the choice of parameters shown in Table \ref{tabA} for the dilaton potential in Eq. \eqref{eq:Vphi}. This particular choice has been used previously in Ref. \cite{Finazzo:2014zga} to calculate the Polyakov loop and the Debye screening mass in a system with a first order confinement-deconfinement phase transition with its thermodynamics closely resembling lattice data for pure $SU(3)$ Yang-Mills theory (see \cite{Finazzo:2014zga} for further details).

We restrict our calculations in this section to temperatures at or above the critical temperature (to be discussed below), which is the range where the black hole solutions constitute the dominant saddle points of the action \eqref{eq:EDaction} in a system with a first order phase transition, in which case we can apply Eq. \eqref{eq:GammaCS} to calculate the Chern-Simons diffusion rate in the high temperature phase.

In order to study the thermodynamics of this system, we follow the standard analysis presented, for example, in Refs. \cite{Gubser:2008ny,Finazzo:2014zga,He:2013qq}. The thermodynamics of the system is characterized by its equation of state, which may be expressed as the temperature dependence of the entropy density $s(T)$, the pressure $p(T)$, the speed of sound squared $c_s^2(T)$, the internal energy density $\epsilon(T)$, or the trace anomaly $I(T)$. In the present holographic setting, one first numerically solves the master equation \eqref{eq:master} and obtains $T$ as a function of the black hole horizon $\phi_H$ using Hawking's formula for the black hole temperature in Eq. \eqref{eq:TphiH}. The entropy density as a function of $\phi_H$ is given by Bekenstein-Hawking's relation \cite{Bekenstein:1973ur,Hawking:1974sw} in Eq. \eqref{eq:sphiH}. Then, $s(T)$ is obtained by numerically interpolating a table of points in the format $\{T(\phi_H),s(\phi_H)\}$ with $\phi_H$ running through different values in the thermodynamically favored range for black hole solutions (to be discussed below). Once one has holographically derived the equation of state $s(T)$, one can obtain $p(T)$, $c_s^2(T)$, $\epsilon(T)$, and $I(T)$ by using standard thermodynamic relations.

\begin{figure}[htp!]
\center
\subfigure[]{\includegraphics[width=0.8\linewidth]{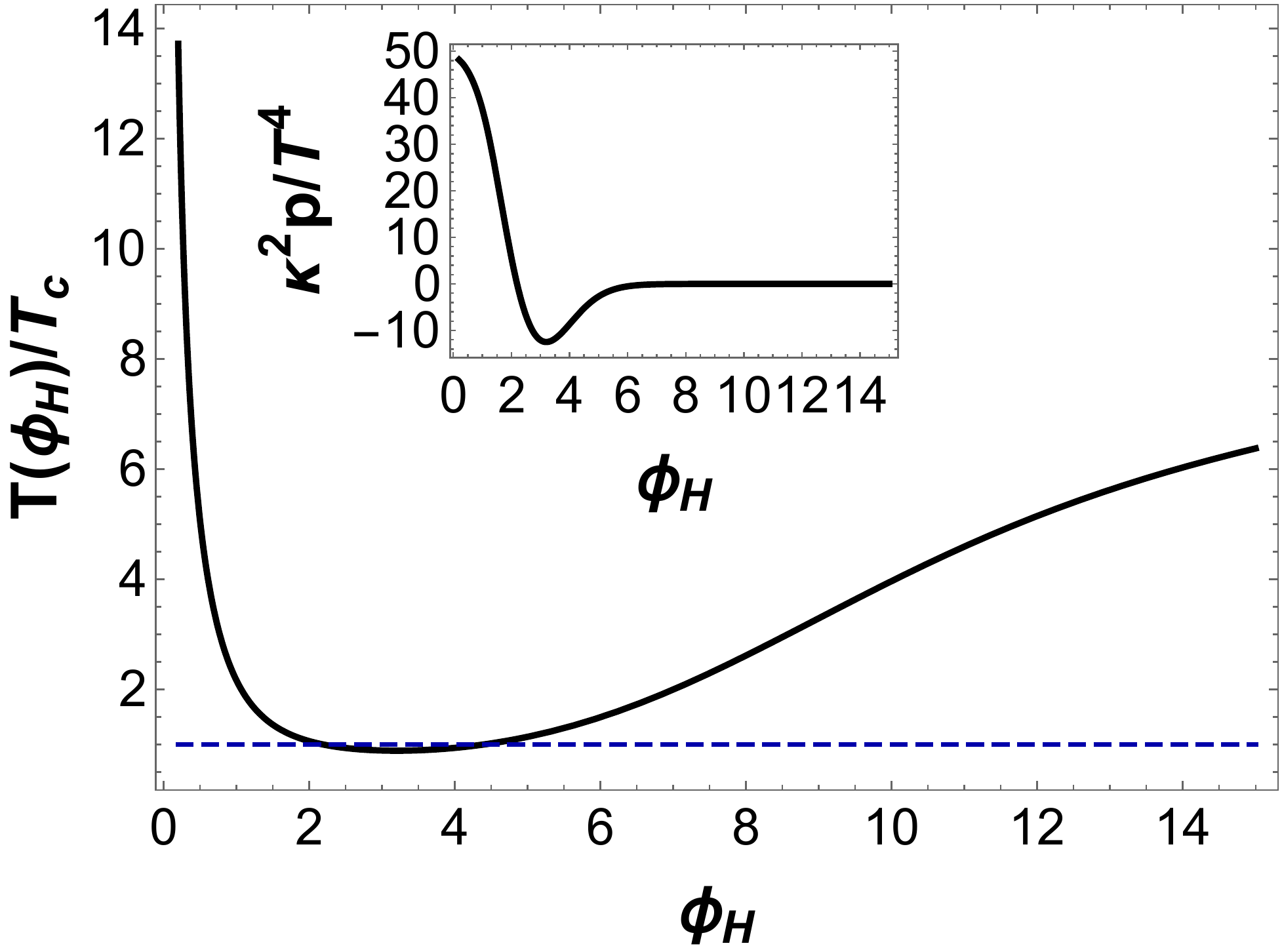}}
\qquad
\subfigure[]{\includegraphics[width=0.8\linewidth]{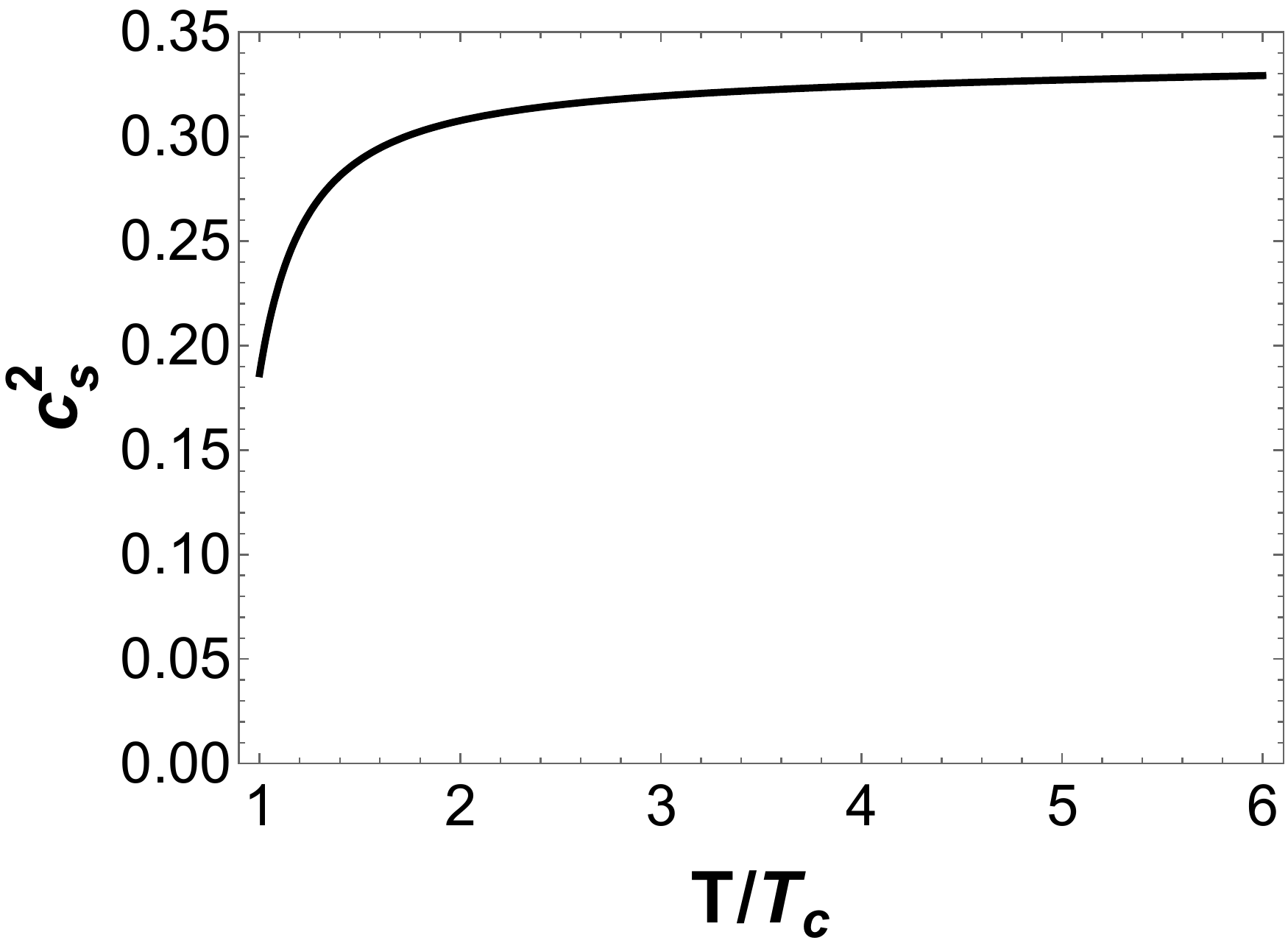}}
\qquad
\subfigure[]{\includegraphics[width=0.8\linewidth]{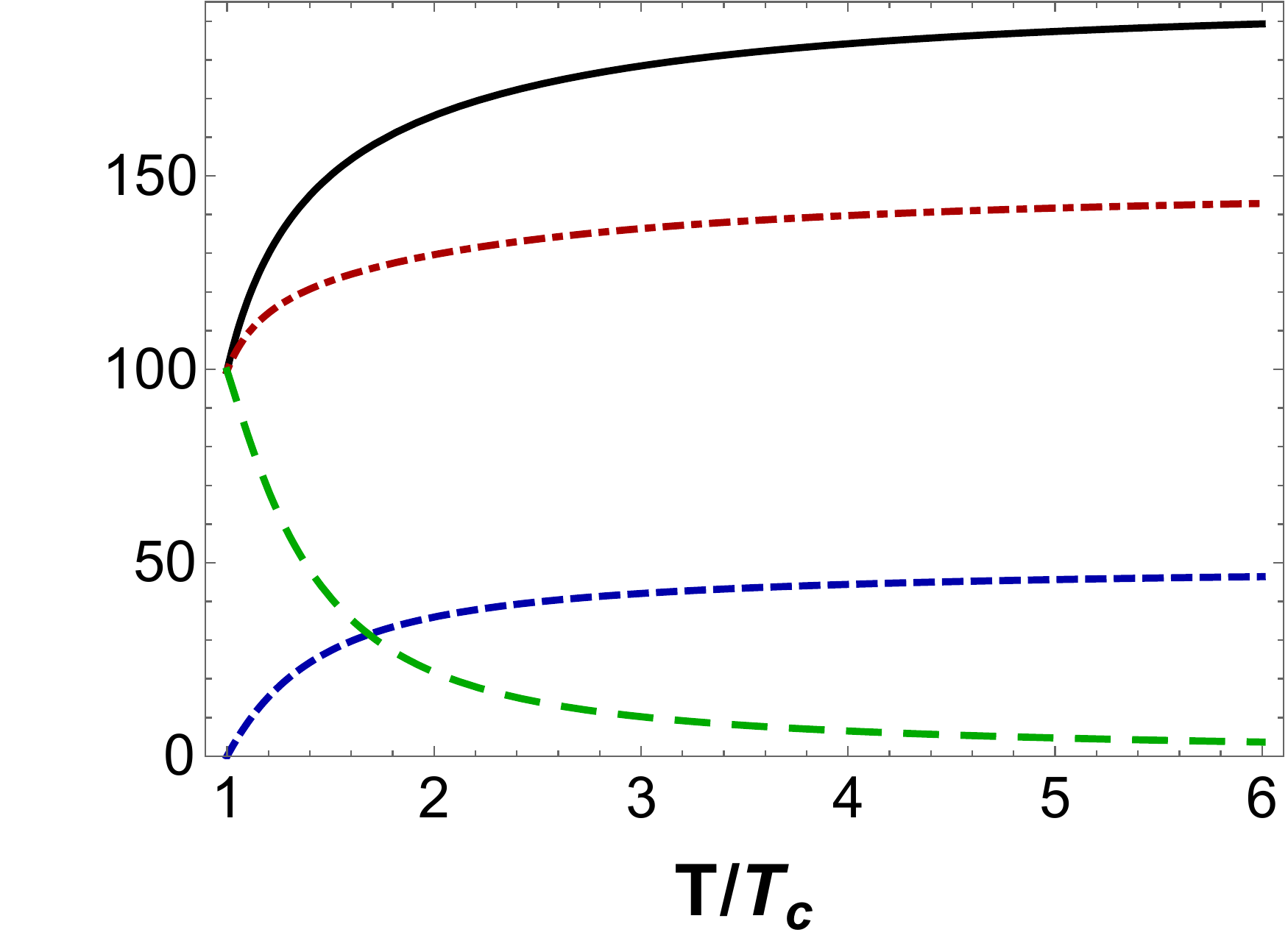}}
\caption{(Color online) Thermodynamics of the Einstein-dilaton model with a 1st order phase transition. (a) Black hole temperature normalized by the critical temperature $T_c$ as a function of the radial position of the black hole horizon $\phi_H$. The dashed line lies at 1, while the minimum temperature for the existence of black hole solutions at the bottom of the curve is $T_{\textrm{min}}\approx 0.9 T_c$. In the inset, we show the range of $\phi_H$ for which the black hole solutions are unstable, displaying negative pressure (within this region, the thermal gas solution corresponds to the dominant saddle point of the model). (b) Speed of sound squared. (c) From top to bottom at high $T$: $\kappa^2 s/T^3$ (solid black curve), $\kappa^2 \epsilon/T^4$ (dot-dashed red curve), $\kappa^2 p/T^4$ (dashed blue curve), and $\kappa^2 I/T^4$ (long-dashed green curve).}
\label{fig:thermoA}
\end{figure}

In Fig. \ref{fig:thermoA} (a), we plot $T(\phi_H)$ normalized by the critical temperature $T(\phi_H^c\approx 2.19)=T_c$, which in a system with a first order phase transition is identified with the temperature at which the black hole pressure (to be discussed below) vanishes coming from positive values at higher temperatures. One notes that $T(\phi_H)/T_c$ has a global minimum at $T(\phi_H^{\textrm{min}}\approx 3.2)=T_{\textrm{min}}\approx 0.9 T_c$, below which there are no black hole solutions. The non-monotonicity of $T(\phi_H)/T_c$ indicates the presence of two competing branches of black hole solutions for $T>T_{\textrm{min}}$, the so-called big black hole solutions identified by the black hole branch with $\phi_H<\phi_H^{\textrm{min}}$, and the small black hole branch identified by $\phi_H>\phi_H^{\textrm{min}}$.\footnote{Note that the larger the value of $\phi_H$ the more distant is the black hole horizon from the boundary of the asymptotically AdS$_5$ space. Therefore, large $\phi_H$ implies a small sized horizon and vice-versa.} There is also a solution to the equations of motion with no horizon, obtained by setting to zero the blackening function $h$ in the metric ansatz, which is the vacuum solution. By going to Euclidean space and compactifying the imaginary Euclidean time direction over a circle with circumference $\beta=1/T$, the vacuum solution is promoted to a thermal gas solution, which exists for any value of $T$. Below $T_{\textrm{min}}$, the thermal gas solution is the only nontrivial solution to the equations of motion (given our ansatze for the Einstein-dilaton fields). However, for $T>T_{\textrm{min}}$ there are three competing solutions, the big and small black hole solutions and the thermal gas solution, and in order to decide which solution is the dominant saddle point of the model at each value of $T$, one must compare their pressures and check which one has the largest pressure: this is the thermodynamically favored solution at a given $T$.

The pressure of the black hole solutions is obtained by integrating the black hole entropy density with respect to the temperature,
\begin{align}
p(T)=\int dT s(T).
\label{eq:pint}
\end{align}
The integration constant in Eq. \eqref{eq:pint} may be fixed by going into the small black hole branch in the limit of vanishing horizon, $\phi_H\rightarrow\infty$, which coincides with the vacuum solution metric, and imposing that in this limit the black hole pressure matches the pressure of the thermal gas background, which is a constant that can be chosen to be zero. Therefore, we may write the black hole pressure as
\begin{align}
p(\phi_H)=\int_\infty^{\phi_H} d\phi_H' \frac{dT(\phi_H')}{d\phi_H'} s(\phi_H').
\label{eq:p(phiH)}
\end{align}
The result is shown in the inset of Fig. \ref{fig:thermoA} (a), where we plotted the dimensionless ratio\footnote{In $\mathcal{N}=4$ Super Yang-Mills theory, one has the following relation between the gravitational constant and the number of colors of the gauge theory: $\kappa^2=4\pi^2L^3/N_c^2$ \cite{Gubser:1996de}. In bottom-up models, as the present one, the precise relation between $\kappa^2$ and $N_c$ is unknown. However, in phenomenological approaches, $\kappa^2$ may be fixed, for instance, by seeding the bottom-up model with lattice QCD data, as done in Refs. \cite{Finazzo:2014zga,Finazzo:2014cna,Rougemont:2015oea,Rougemont:2015wca,Rougemont:2015ona,Finazzo:2015xwa}. In the present work, we do not worry to fix any value for $\kappa^2$ since this will be of no relevance for the qualitative results we are going to obtain (note that the only thing $\kappa^2$ can do is to shift the height of the curves without changing their shapes).} $\kappa^2 p/T^4$. One can see that for $\phi_H<\phi_H^c$, the black hole pressure is positive, going to zero at $\phi_H=\phi_H^c$, while it becomes negative for $\phi_H>\phi_H^c$, asymptoting to zero from below for $\phi_H\rightarrow\infty$. Therefore, one concludes that the big black hole branch is thermodynamically favored for $T>T_c$, while below $T_c$ the thermal gas solution is the dominant saddle point. Consequently, at $T=T_c$ the system undergoes a first order Hawking-Page phase transition from black hole geometries to thermal gas geometries \cite{Hawking:1982dh}. Note also that the small black hole branch is always thermodynamically unstable in the present model.

One may obtain $p(T)$ by numerically interpolating a table of points in the format $\{T(\phi_H),p(\phi_H)\}$ with $\phi_H$ running from\footnote{Corresponding to the end point of the stable branch of big black hole solutions. For $\phi_H>\phi_H^c$ the pressure of the system is zero, corresponding to the pressure of the thermal gas solution, as discussed before.} $\phi_H^c$ up to some small numeric value corresponding to high $T$ geometries\footnote{For $T\gg T_c$, the pressure and all physical observables of the model tend to their conformal values.}. The internal energy density is given by $\epsilon(T)=Ts(T)-p(T)$, while the trace anomaly is defined as $I(T)=\epsilon(T)-3p(T)$.\footnote{In conformal field theories as $\mathcal{N}=4$ Super Yang-Mills, the trace anomaly vanishes.} The speed of sound squared is obtained through the relation $c_s^2(T)=dp/d\epsilon=d\ln(T)/d\ln(s)$. The corresponding results for dimensionless ratios of these thermodynamic variables normalized by adequate powers of $T$ are shown in Fig. \ref{fig:thermoA} (b) and (c). One observes that $\kappa^2s/T^3$, $c_s^2$, $\kappa^2\epsilon/T^4$, and $\kappa^2I/T^4$ jump at $T=T_c$, while the pressure vanishes. For $T\gg T_c$, all these dimensionless observables tend to their conformal values. In particular, $\kappa^2I/T^4$ goes to zero at high temperatures, when our model recovers conformal invariance.

\begin{figure}[htp!]
\center
\includegraphics[width=0.45\textwidth]{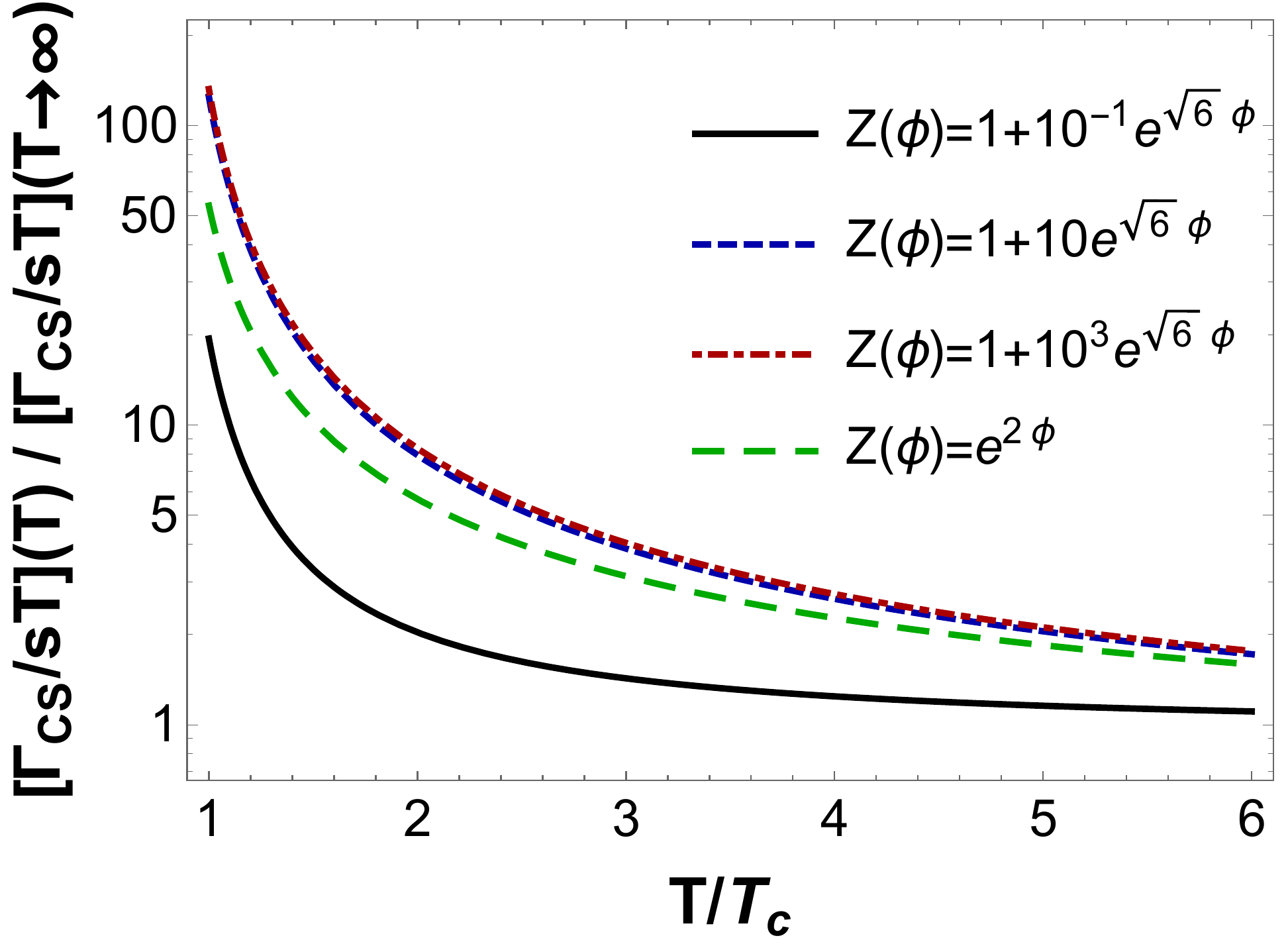}
\caption{(Color online) Ratio of the Chern-Simons diffusion rate divided by the product $sT$ normalized by its value at large $T$ in the Einstein-dilaton model with a 1st order phase transition for different axion-dilaton coupling functions.}
\label{fig:CSdiffA}
\end{figure}

In Fig. \ref{fig:CSdiffA}, we show our results for the dimensionless ratio $\Gamma_{\textrm{CS}}/sT$ normalized by its value at $T\rightarrow\infty$ for different choices of the axion-dilaton coupling function. One notes that the different axion-dilaton couplings produce the same qualitative results, with the Chern-Simons diffusion rate jumping at $T_c$ in a first order phase transition \cite{Gursoy:2012bt}. The fact that the different profiles for the axion-dilaton coupling in Eqs. \eqref{eq:ZKiritsis} and \eqref{eq:ZTrancanelli} give the same qualitative results is mainly related to the fact that both couplings present an exponential dependence on $\phi$, therefore, from Eq. \eqref{eq:GammaCS}, $\Gamma_{\textrm{CS}}/sT$ depends mainly on $e^{\textrm{cte}\times\phi_H(T)}$. Since $\phi_H(T)$ is determined by the chosen profile for $V(\phi)$, the different axion-dilaton couplings in Eqs. \eqref{eq:ZKiritsis} and \eqref{eq:ZTrancanelli} can only produce quantitatively distinct results, with the qualitative features remaining the same. Therefore, as we are going to see in the next two sections, the qualitative behavior of $\Gamma_{\textrm{CS}}/sT$ will only change by modifying the nature of the phase transition encoded in the choice of parameters of the dilaton potential in Eq. \eqref{eq:Vphi}.

It is also interesting to note the saturating behavior observed in Fig. \ref{fig:CSdiffA} by using the coupling in Eq. \eqref{eq:ZKiritsis} with $Z_1>10$. This same kind of saturation will also hold in the cases of a second order phase transition and a crossover, to be discussed in the sequel.

\section{Model B: second order phase transition}

\begin{table}[h]
 \begin{center}
  \begin{tabular}{| c | c | c | c | c |}
    \hline
    $a$ & $\gamma$ & $b_2$ & $b_4$ & $b_6$ \\
    \hline
    \hline
    0 & $\sqrt{2}/2$ & $1.942$ & $0$ & $0$ \\
    \hline
  \end{tabular}
 \caption{A choice of parameters of the dilaton potential in Eq. \eqref{eq:Vphi} giving a system with a second order phase transition. \label{tabB}}
 \end{center}
\end{table}

In this section, we discuss the thermodynamics and the behavior of the Chern-Simons diffusion rate in a system with a second order phase transition, which may be modeled by the choice of parameters shown in Table \ref{tabB} for the dilaton potential in Eq. \eqref{eq:Vphi}. This particular choice has been used previously in Ref. \cite{Gubser:2008ny}.

\begin{figure}[htp!]
\center
\subfigure[]{\includegraphics[width=0.8\linewidth]{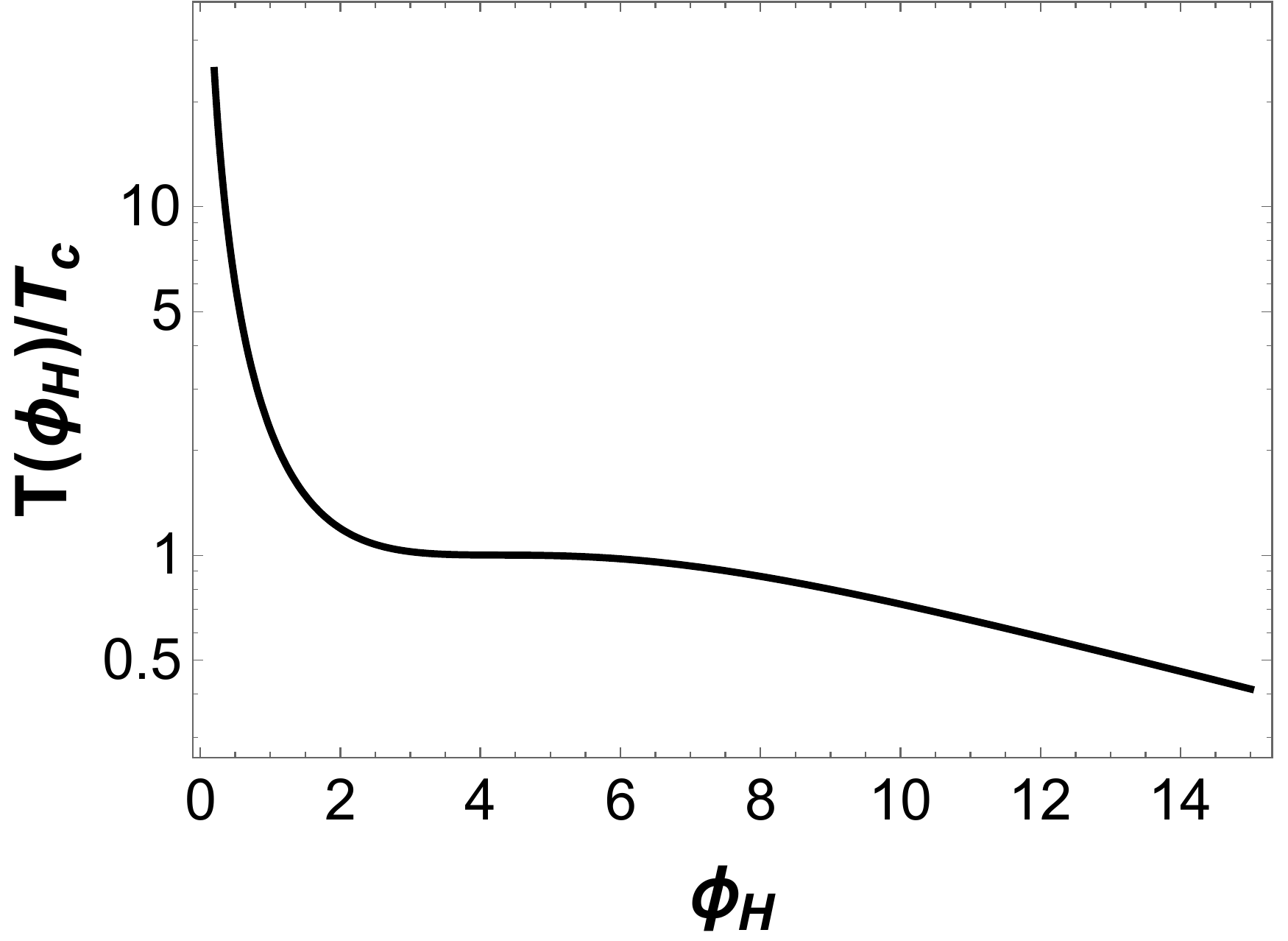}}
\qquad
\subfigure[]{\includegraphics[width=0.8\linewidth]{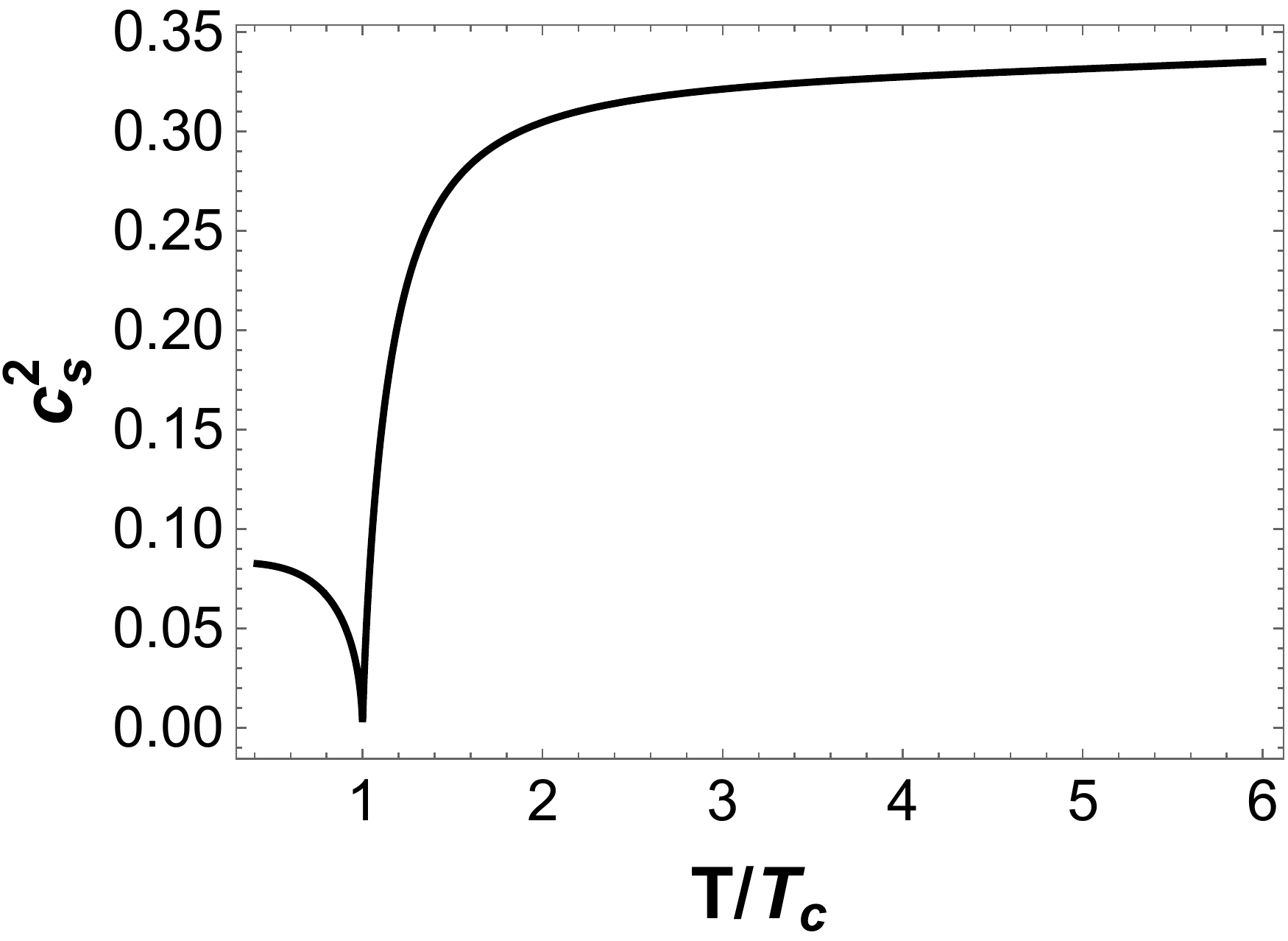}}
\qquad
\subfigure[]{\includegraphics[width=0.8\linewidth]{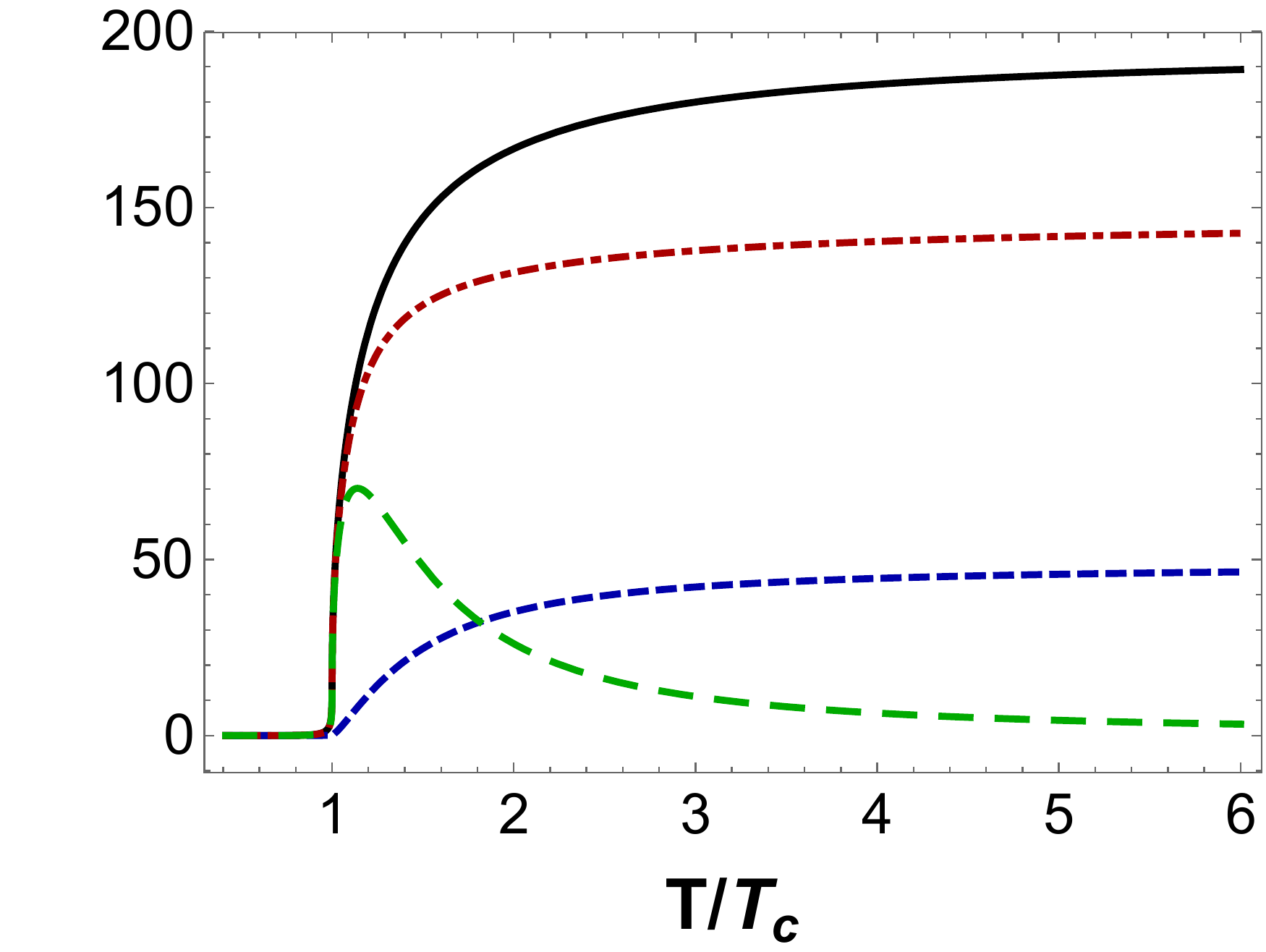}}
\caption{(Color online) Thermodynamics of the Einstein-dilaton model with a 2nd order phase transition. (a) Black hole temperature normalized by the critical temperature $T_c$ as a function of the radial position of the black hole horizon $\phi_H$. At the critical temperature the curve has a stationary point of inflection. (b) Speed of sound squared: it is never negative, but it vanishes at $T=T_c$. (c) From top to bottom at high $T$: $\kappa^2 s/T^3$ (solid black curve), $\kappa^2 \epsilon/T^4$ (dot-dashed red curve), $\kappa^2 p/T^4$ (dashed blue curve), and $\kappa^2 I/T^4$ (long-dashed green curve).}
\label{fig:thermoB}
\end{figure}

In a system with a second order phase transition, we identify the critical temperature $T(\phi_H^c\approx 4.26)=T_c$ as the point where $c_s^2$ vanishes (although it must be non-negative) \cite{Gubser:2008ny}. In Fig. \ref{fig:thermoB} (a), one notes that $\phi_H=\phi_H^c$ is a stationary inflection point of $T(\phi_H)/T_c$. In Fig. \ref{fig:thermoB} (b) and (c) we plot the equation of state. One can see that at $T=T_c$, $s/T^3$ develops an infinite slope and its temperature derivative diverges, as usual in second order phase transitions. Also, the trace anomaly peaks at $T=T_c$, while going to zero at high temperatures (corresponding to the conformal limit).

In the present case, contrary to what happens in a system with a first order Hawking-Page phase transition, the black hole solutions exist for any value of $T$ and are always thermodynamically favored over the thermal gas solution (which has zero pressure).

\begin{figure}[htp!]
\center
\includegraphics[width=0.45\textwidth]{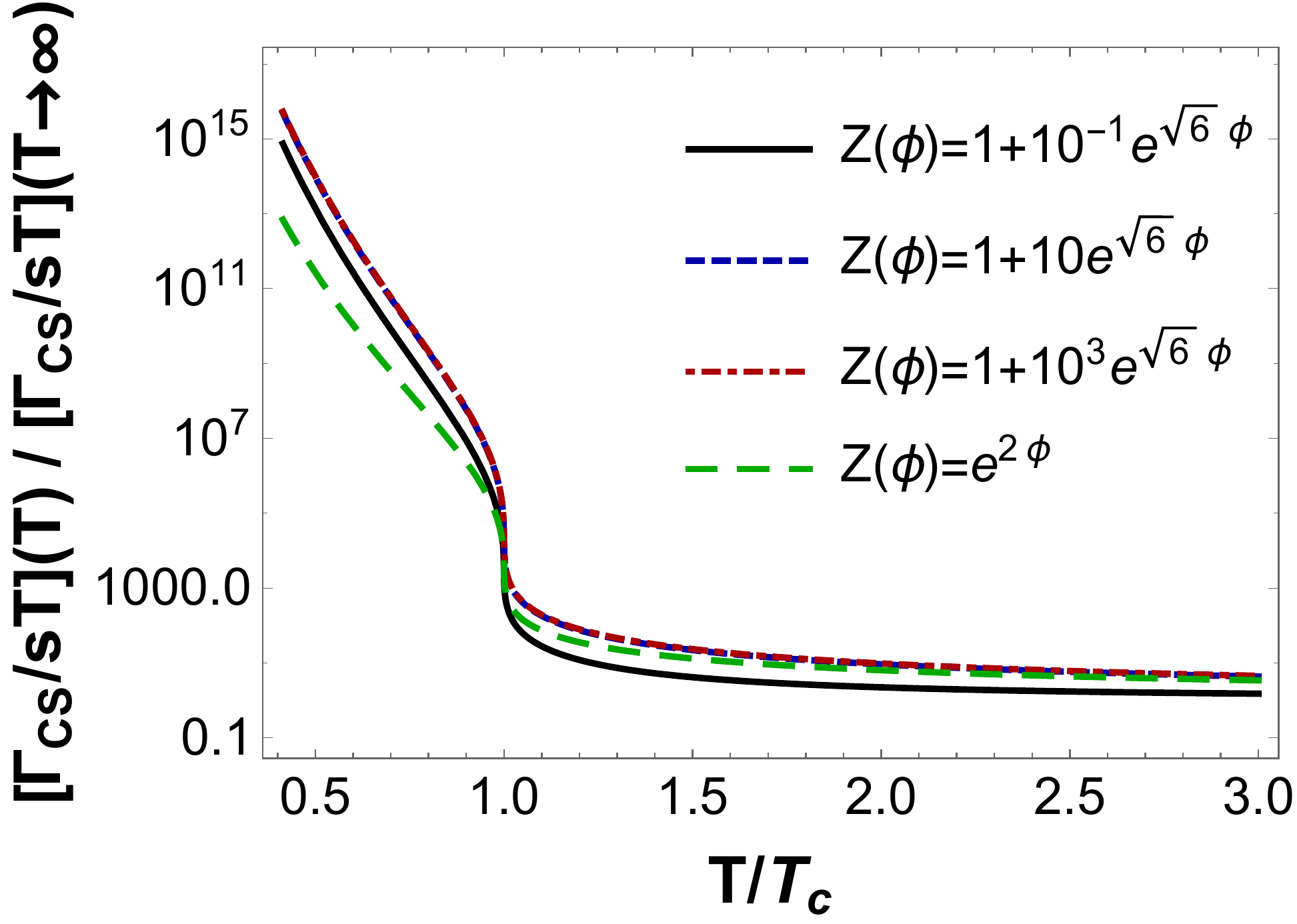}
\caption{(Color online) Ratio of the Chern-Simons diffusion rate divided by the product $sT$ normalized by its value at large $T$ in the Einstein-dilaton model with a 2nd order phase transition for different axion-dilaton coupling functions.}
\label{fig:CSdiffB}
\end{figure}

In Fig. \ref{fig:CSdiffB}, we display our results for $\Gamma_{\textrm{CS}}/sT$ normalized by its value at $T\rightarrow\infty$ for different choices of the axion-dilaton coupling function. One notes that $\Gamma_{\textrm{CS}}/sT$ has infinite slope at $T=T_c$ in a second order phase transition. Moreover, the Chern-Simons diffusion rate grows by orders of magnitude below the critical temperature, which is a consequence of the exponential behavior of $\Gamma_{\textrm{CS}}/sT$ in conjunction with the inverse relation $\phi_H(T)$, which may be inferred from the plot in Fig. \ref{fig:thermoB} (a).

\section{Model C: crossover}

\begin{table}[h]
 \begin{center}
  \begin{tabular}{| c | c | c | c | c |}
    \hline
    $a$ & $\gamma$ & $b_2$ & $b_4$ & $b_6$ \\
    \hline
    \hline
    0 & $0.606$ & $2.06$ & $0$ & $0$ \\
    \hline
  \end{tabular}
 \caption{A choice of parameters of the dilaton potential in Eq. \eqref{eq:Vphi} yielding a system with a crossover. \label{tabC}}
 \end{center}
\end{table}

In this section, we discuss the thermodynamics and the behavior of the Chern-Simons diffusion rate in a system with a crossover, which may be modeled by the choice of parameters shown in Table \ref{tabC} for the dilaton potential in Eq. \eqref{eq:Vphi}. This particular choice has been used previously in Refs. \cite{Gubser:2008ny,DeWolfe:2010he,DeWolfe:2011ts} to holographically emulate the thermodynamic properties of $SU(3)$ $(2+1)$-flavor QCD.

\begin{figure}[htp!]
\center
\subfigure[]{\includegraphics[width=0.8\linewidth]{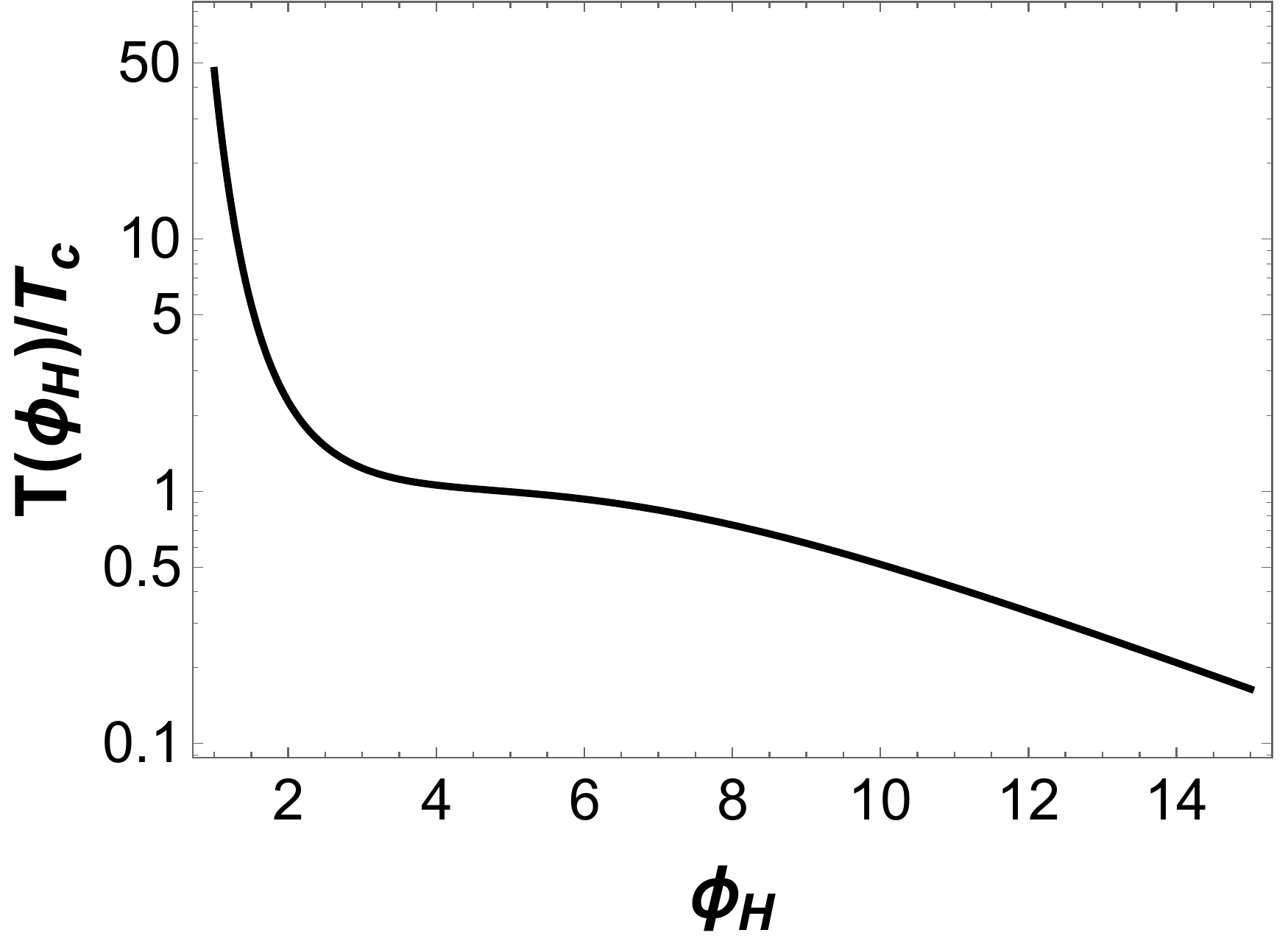}}
\qquad
\subfigure[]{\includegraphics[width=0.8\linewidth]{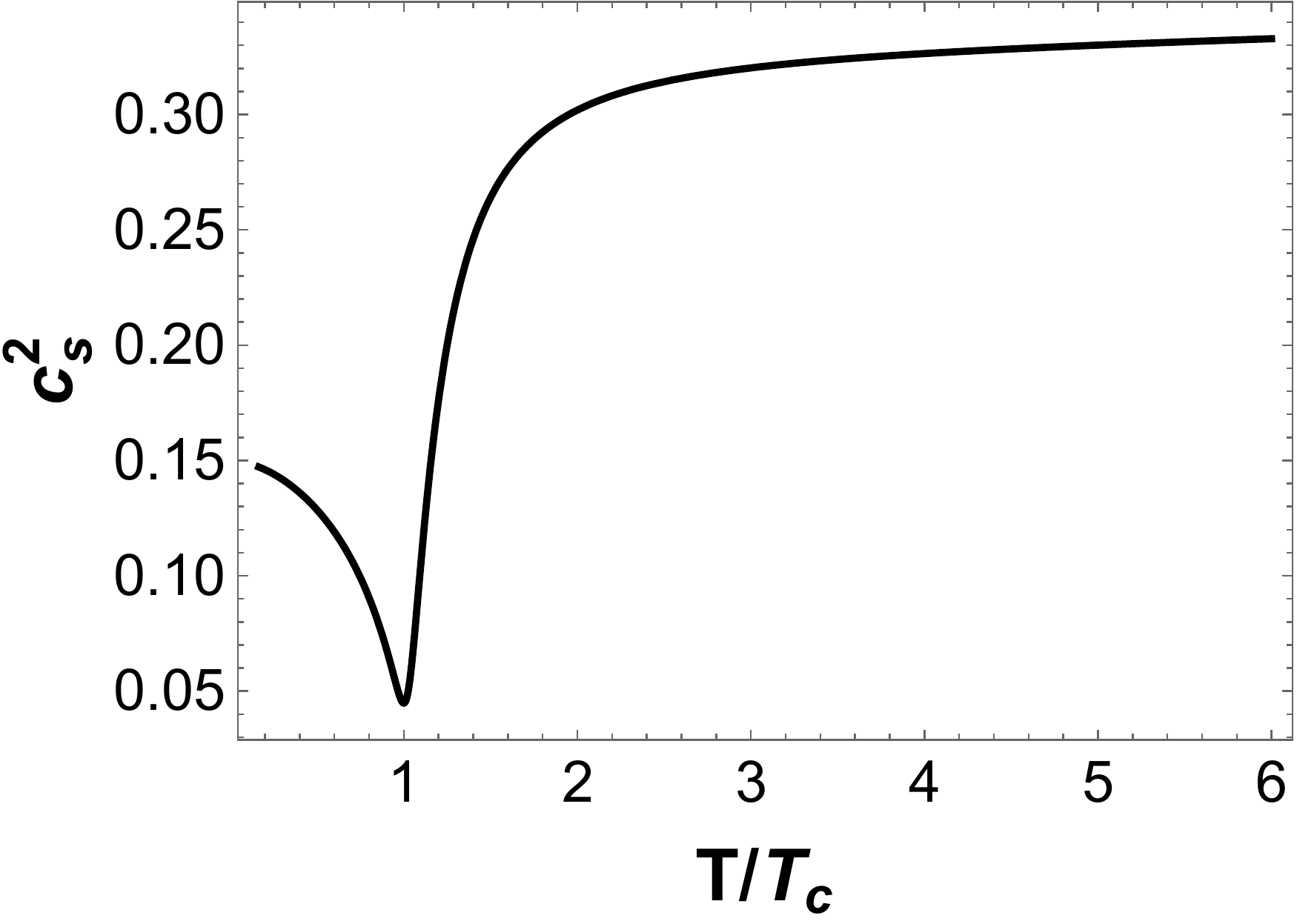}}
\qquad
\subfigure[]{\includegraphics[width=0.8\linewidth]{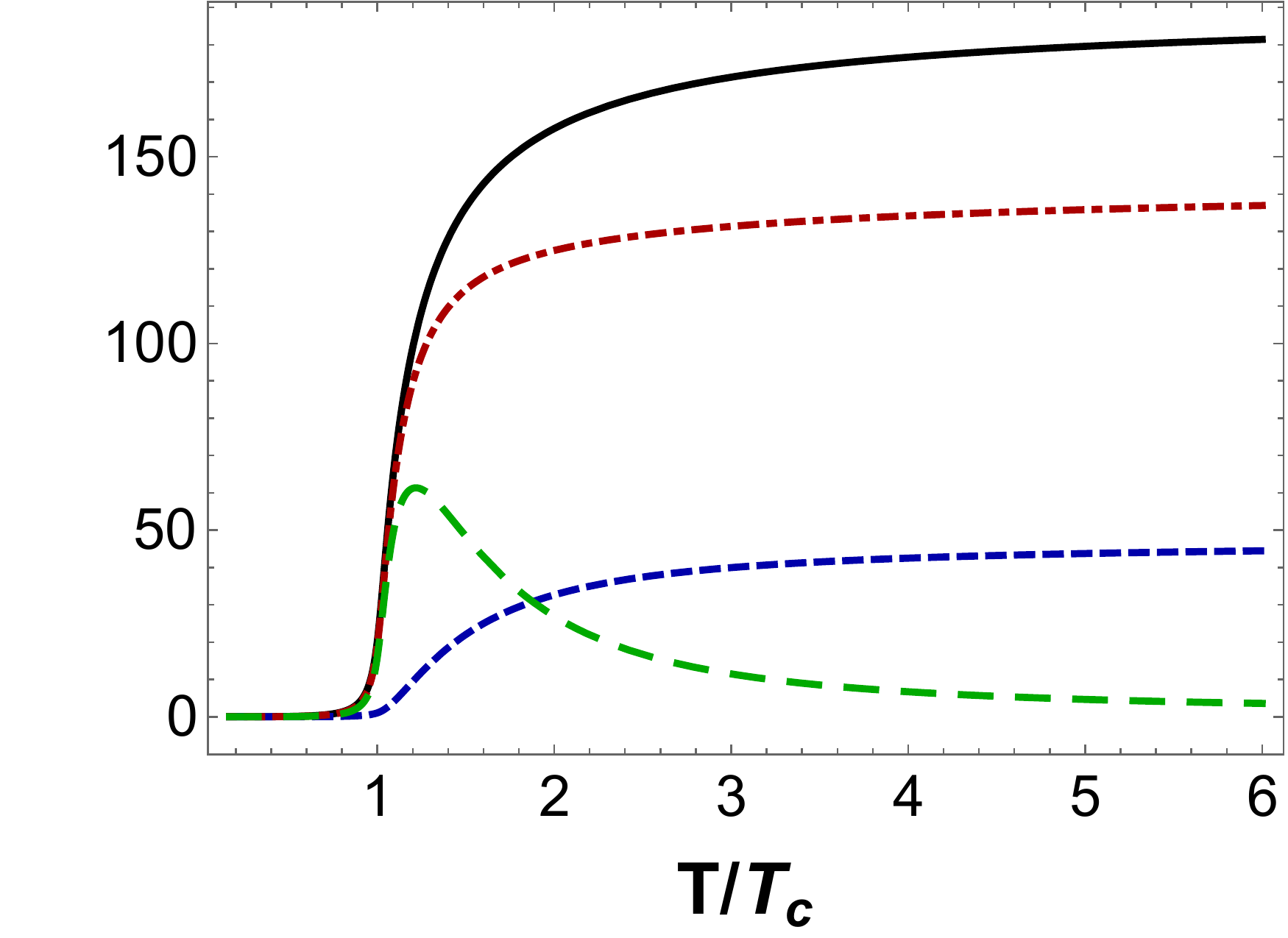}}
\caption{(Color online) Thermodynamics of the Einstein-dilaton model with a crossover. (a) Black hole temperature normalized by the critical temperature $T_c$ as a function of the radial position of the black hole horizon $\phi_H$. (b) Speed of sound squared. (c) From top to bottom at high $T$: $\kappa^2 s/T^3$ (solid black curve), $\kappa^2 \epsilon/T^4$ (dot-dashed red curve), $\kappa^2 p/T^4$ (dashed blue curve), and $\kappa^2 I/T^4$ (long-dashed green curve).}
\label{fig:thermoC}
\end{figure}

In the case of a crossover, the Helmholtz free energy density does not possess any non-analyticity, contrary to what happens in a real phase transition (for instance, in the first order system analyzed before, the first order temperature derivative of the free energy jumps at $T=T_c$, while in the second order system, the second order derivative diverges). Consequently, all its derivatives are smooth and the definition of the so-called pseudo-critical temperature (which we shall denote also as ``$T_c$'') in this case is not unique and depends on the thermodynamic observable employed to characterize it \cite{Borsanyi:2010bp}. In a crossover, even though there is no real phase transition, the degrees of freedom of the system rapidly change across a small temperature interval, and some possible benchmarks\footnote{More precisely, these different characteristic points may be used to define a band for the crossover temperature.} used to characterize the pseudo-critical temperature are, for instance, the inflection point of $s/T^3$, the peak of $I/T^4$, or the minimum of $c_s^2$. Here we choose the latter one to characterize $T_c$.

In Fig. \ref{fig:thermoC} (a), we plot $T(\phi_H)/T_c$ in the system with a crossover. One notes that, as in the system with a second order phase transition, the black hole solutions exist for any value of $T$, but this time the inflection point of $T(\phi_H)/T_c$ is non-stationary. In Fig. \ref{fig:thermoC} (b) and (c) we plot the equation of state, whose behavior in a crossover looks like a ``smoothed out version of the equation of state in the system with a second order phase transition'', with the main differences being that $c_s^2$ no longer vanishes at $T=T_c$ and $s/T^3$ has finite slope, although rapidly varying around $T=T_c$. Also in this case, the black hole solutions are always thermodynamically favored over the zero pressure thermal gas solution.

\begin{figure}[htp!]
\center
\includegraphics[width=0.45\textwidth]{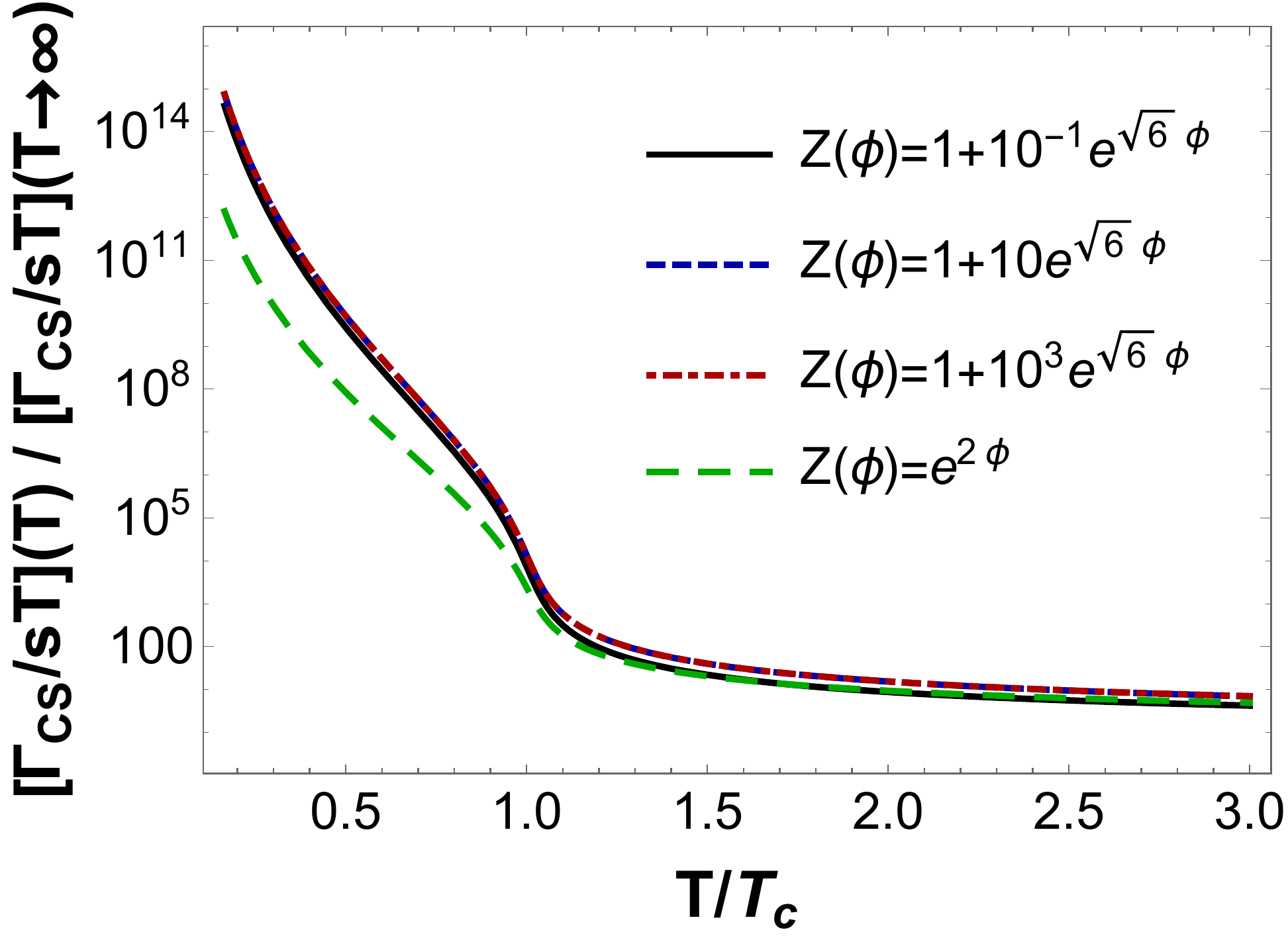}
\caption{(Color online) Ratio of the Chern-Simons diffusion rate divided by the product $sT$ normalized by its value at large $T$ in the Einstein-dilaton model with a crossover for different axion-dilaton coupling functions.}
\label{fig:CSdiffC}
\end{figure}

In Fig. \ref{fig:CSdiffC}, we present our results for $\Gamma_{\textrm{CS}}/sT$ normalized by its value at $T\rightarrow\infty$ for different choices of the axion-dilaton coupling function. One notes that $\Gamma_{\textrm{CS}}/sT$ behaves smoothly,
although displaying a fast variation around the pseudo-critical temperature. As in the case of a second order phase transition, the Chern-Simons diffusion rate grows by orders of magnitude below $T_c$ in a crossover.

\section{Further discussion}

In this section, we discuss in depth the limitations of the holographic results obtained in the previous sections when regarding their possible applicability to real world QCD phenomenology. The main point we want to stress is that the approach followed in the present work is qualitative rather than quantitative, therefore, the main results obtained here refer to the qualitative behavior of the Chern-Simons diffusion rate close to different phase transitions.

The calculations carried out here took place in the holographic arena, where one describes the physics of a strongly coupled gauge theory with a large number of colors in flat spacetime using classical gravity in curved spacetimes with higher dimension. Therefore, one cannot expect that the results of any gauge/gravity model defined in the classical gravity limit of the holographic correspondence would apply to describe the perturbative limit of QCD setting up at high temperatures (when compared to the $\Lambda_{\textrm{QCD}}$ scale). In fact, in the classical gravity limit, the dual gauge theory is always strongly coupled, having a non-trivial ultraviolet fixed point, differently from the trivial ultraviolet fixed point of QCD, which describes the perturbative phenomenon of asymptotic freedom. Moreover, the classical gravity limit of the holographic correspondence cannot describe also the thermodynamics of the confined hadron gas phase, where the pressure goes like $\sim N_c^0$, instead of $\sim N_c^2$ as in the deconfined plasma phase. Therefore, the limits of low and high temperatures in the holographic settings worked out in the present work should not be expected to give a reliable description of real world QCD phenomenology.

On the other hand, around the crossover transition in QCD at zero chemical potential (or also the first order phase transition in pure Yang-Mills theory), and possibly also around the long-sought QCD CEP at nonzero chemical potential (where first order, second order and crossover transitions come into play), QCD is in the strongly coupled regime, where perturbative techniques cannot be reliably applied. Moreover, when referring to non-equilibrium, real time observables, like the Chern-Simons diffusion rate, lattice QCD methods still face severe technical difficulties. Consequently, at present, nothing is known from first principle calculations about the behavior of the Chern-Simons diffusion rate close to these different phase transitions. And it is precisely in this regime of strong coupling where holographic techniques may be useful to give at least qualitative insights on the behavior of both, equilibrium and also non-equilibrium, real time observables.

One may also speculate another point of concern in terms of the possible applicability of our findings to real world QCD phenomenology, in relation with the Witten-Veneziano formula \cite{Witten:1979vv,Veneziano:1979ec}. According to this formula, in the large $N_c$ limit the mass of the pseudoscalar meson $\eta'$ should remain light due to the $N_c$ suppression of the axial anomaly, which is different from real world QCD, where $\eta'$ is significantly heavier than the octet of pseudoscalar mesons $(\pi^+,\pi^-,\pi^0,K^+,K^-,K^0,\bar{K}^0,\eta)$ associated with the chiral symmetry breaking. The expectation in holographic models is that the $\eta'$ mass should be lifted as one includes $1/N_c$ corrections in the gauge/gravity duality. A hint that this is so is suggested by Ref. \cite{Barbon:2004dq}, which identified, in the context of the Sakai-Sugimoto model \cite{Sakai:2004cn} with $N_f$ flavor D6-branes, the relevant stringy corrections and estimated the correct $N_f/N_c$ dependence of the $\eta'$ mass, in accordance with the Witten-Veneziano formula. However, the Witten-Veneziano formula depends on topological properties giving rise to the vacuum topological susceptibility $\chi_{\textrm{top}}$. It has been argued since the first works on the sphaleron diffusion rate \cite{Arnold:1987zg} that there is a certain independence between the real time retarded correlation function (which describes $\Gamma_{\textrm{CS}}$) and the Euclidean correlator (which yields $\chi_{\textrm{top}}$). Consequently, the lightness of $\eta'$ in the large $N_c$ limit cannot be used to infer the fate of the Chern-Simons diffusion rate.

Keeping in mind the caveats and limitations discussed above, we expect that the main qualitative results of the present work, namely, that the ratio $\Gamma_{\textrm{CS}}/sT$ either jumps, acquires an infinite slope, or behaves smoothly, although displaying a fast variation around $T_c$, depending on the phase transition being either of first order, second order, or a crossover, respectively, could also apply to real world QCD.

All in all, our findings in the present work add to the literature the first predictions for the Chern-Simons diffusion rate across second order and crossover transitions in strongly coupled non-conformal, non-Abelian gauge theories.

\section{Concluding remarks}

In the present work, we conducted a systematic study of the behavior of the Chern-Simons diffusion rate across different kinds of phase transitions. By using a bottom-up Einstein-dilaton holographic setup, we adjusted the dilaton potential in order to obtain different equations of state in the gauge theory, displaying a first order, a second order, or a crossover transition. As originally found in Ref. \cite{Gursoy:2012bt}, we concluded that in the case of a first order phase transition the dimensionless ratio $\Gamma_{\textrm{CS}}/sT$ jumps at the critical temperature. The new conclusions obtained in the present work regard the behavior of $\Gamma_{\textrm{CS}}/sT$ across a second order phase transition, where it develops an infinite slope at the critical temperature, and also its behavior through a crossover, where it behaves smoothly, although displaying a fast variation around the pseudo-critical temperature. In all the cases, $\Gamma_{\textrm{CS}}/sT$ increases with decreasing temperature.

Our results give the first predictions for the Chern-Simons diffusion rate across second order and crossover transitions in strongly coupled non-conformal, non-Abelian gauge theories. Moreover, our conclusions were shown to be robust in face of the choice of different profiles for the axion-dilaton coupling previously derived in the literature.

The behavior of the Chern-Simons diffusion rate across these different phase transitions is expected to play a relevant role for the CME around the QCD critical end point, which is a second order phase transition point connecting a crossover band to a line of first order phase transition.

It will be also interesting to investigate the behavior of the spectral density and the quasinormal modes in the axion channel for each of the three different phase transitions discussed here, which is an on-going project.

\begin{acknowledgments}
We thank J. Noronha, V. Jahnke, H. Nastase, and R. Critelli for fruitful discussions. R.R. acknowledges financial support by the S\~{a}o Paulo Research Foundation (FA\-PESP) under FAPESP grant number 2013/04036-0. S.I.F. was supported by FAPESP and Coordena\c{c}\~ao de Aperfei\c{c}oamen\-to de Pessoal de N\'{i}vel Superior (CAPES) under FAPESP grant number 2015/00240-7.
\end{acknowledgments}


\end{document}